\documentclass[aps,prx,showkeys,twocolumn,groupedaddress]{revtex4-1}

\usepackage{graphicx,graphics,color,epsfig}
\usepackage{amsmath}
\usepackage{amssymb}
\usepackage{amsfonts}
\usepackage{amsfonts}
\usepackage{epstopdf}
\usepackage{bm}
\usepackage{times,xspace}
\usepackage{color}

\begin{document}

\title{Theory of Weyl orbital semimetals and predictions of several materials classes}

\author{Kapildeb Dolui$^1$, and Tanmoy Das$^{1,2}$}

\affiliation{$^1$Graphene Research Center and Department of Physics, National University of Singapore, 2 Science Drive 3, 117542, Singapore\\
$^2$Department of Physics, Indian Institute of Physics, Bangalore 560012, India.}
\date{\today}


\begin{abstract}
Graphene, topological insulators, and Weyl semimetals are three widely studied materials classes which possess Dirac or Weyl cones arising from either sublattice symmetry or spin-orbit coupling. In this work, we present a theory of a new class of bulk Dirac and Weyl cones, dubbed Weyl orbital semimetals, where the orbital polarization and texture inversion between two electronic states at discrete momenta lend itself into protected Dirac or Weyl cones without spin-orbit coupling. We also predict several families of Weyl orbital semimetals including V$_3$S$_4$, NiTi$_3$S$_6$, BLi, and PbO$_2$ via first-principle band structure calculations. We find that the highest Fermi velocity predicted in some of these materials is even larger than that of the existing Dirac materials. The synthesis of Weyl orbital semimetals will not only expand the territory of Dirac materials beyond the quintessential spin-orbit coupled systems and hexagonal lattice to the entire periodic table, but it may also open up new possibilities for orbital controlled electronics or `orbitronics'.
\end{abstract}


\maketitle
\section{Introduction} 
Dirac fermions emerge in solid state systems when the band structure is embedded with any two \ component quantum degrees of freedom such as spin or pseudospin under specific symmetry considerations.\cite{CastroNeto,Hasanreview,Zhangreview,WeylreviewBalents,Weylreview,Weylreview2}  In graphene, gapless Dirac cones arise due to the sublattice symmetry of the honeycomb lattice when the sample dimension is reduced exactly to an atomically thin two-dimensional (2D) sheet.\cite{CastroNeto} In topological insulators, an `inverted' bulk bandgap, typically opened by spin-orbit coupling (SOC), renders Dirac excitations on the surface or boundary of the sample as long as time-reversal symmetry is held.\cite{Hasanreview,Zhangreview} Some of these special conditions required for the formation of Dirac fermions in graphene and topological insulators are lifted in the Weyl semimetal framework. In the latter family, Weyl cone is formed in the bulk band structure where bulk conduction and valence bands meet only at discrete momenta, and is protected by lattice symmetry.\cite{WeylreviewBalents,Weylreview,Weylreview2} Because Weyl nodes are easily accessible for room-temperature applications, tremendous research activities have been devoted in recent years for the prediction, discovery, and engineering of new Weyl semimetals families. The predicted materials so far include Iridates,\cite{Iridates}  HgCr$_2$Se$_4$,\cite{HgCrSe,HgCrSe2} $A_3$Bi ($A$=Na, K, Rb),\cite{Na3Bi} $\beta$-cristobaline BiO$_2$,\cite{BiO2} and also in engineered heterostructures\cite{WeylTI,DasWeyl}. Experimentally, Cd$_3$As$_2$ \cite{ExpCdAsBorisenko,ExpCdAsHasan,ExpCdAsAli} and Na$_3$Bi \cite{ExpNa3Bi,ExpNa3BiHasan} have been successfully synthesized to date as bulk Dirac materials.
   
Because SOC is a common ingredient for the formation of Weyl cone in the existing Weyl semimetals and topological insulators, the corresponding materials selection is limited to materials with heavy elements and non-magnetic ground state. However, for spintronics and other transport related purposes, heavy-elements are less effective because they are prone to strong many-body interaction and magnetism which significantly enhance Dirac mass via renormalization and/ or band gapping.\cite{Wray} Some of the other possibilities are pseudospintronics,\cite{pseudospintronics,CastroNeto} and valleytronics~\cite{ValeytronicsBeenakker,ValleytronicsXiao,Valleytronics} where the quantum control of electric current is achieved by the definite polarization of the sublattice and `valley' degrees of freedom, respectively. Here the precise lattice symmetry and structural confinement play the central role to achieve high degree of phase coherence for the electron transport. 

In this paper, we explore a different idea for the formation of bulk Dirac cones and Weyl orbital semimetals without the need of SOC or structural confinement. We develop a generic low-energy theory for the Weyl orbital semimetals for various combinations of different orbitals, such as  even and odd orbitals pair or  bonding and antibonding states or pair of even or pair of odd orbitals or two different basis of same orbital, in variety of 3D lattices. The key ingredient in our theory roots in finding the optimum conditions in which the two orbitals states commence inverted band structure, and at the same time their inter-orbital electron hoppings obtain an odd function of energy-momentum dispersion, such that its low-energy Hamiltonian can be reduced to an effective~${\bf k}\cdot{\bf p}$ - type Hamiltonian. The resulting 3D Dirac or Weyl nodes at the orbital degenerate points are protected by orbital  symmetry. By using density functional theory (DFT), we predict four distinct classes of materials which exemplify different mechanisms for generating Weyl orbital nodes. The predicted materials have orbital components stemming from the weakly correlated $p$ or $d$ orbitals, which, thereby, stipulates very high Weyl fermion velocity. For BLi, our first-principle estimation of Weyl fermion velocity is $\sim 2\times 10^{6}$ m/s, which is even larger than graphene. Finally, we discuss the possibilities of obtaining orbitronics and quantum orbital Hall effect in the Weyl orbital semimetals.

\vskip0.5cm
\noindent
\section{Theory}

\begin{figure}[top]
\rotatebox[origin=c]{0}{\includegraphics[width=0.95\columnwidth]{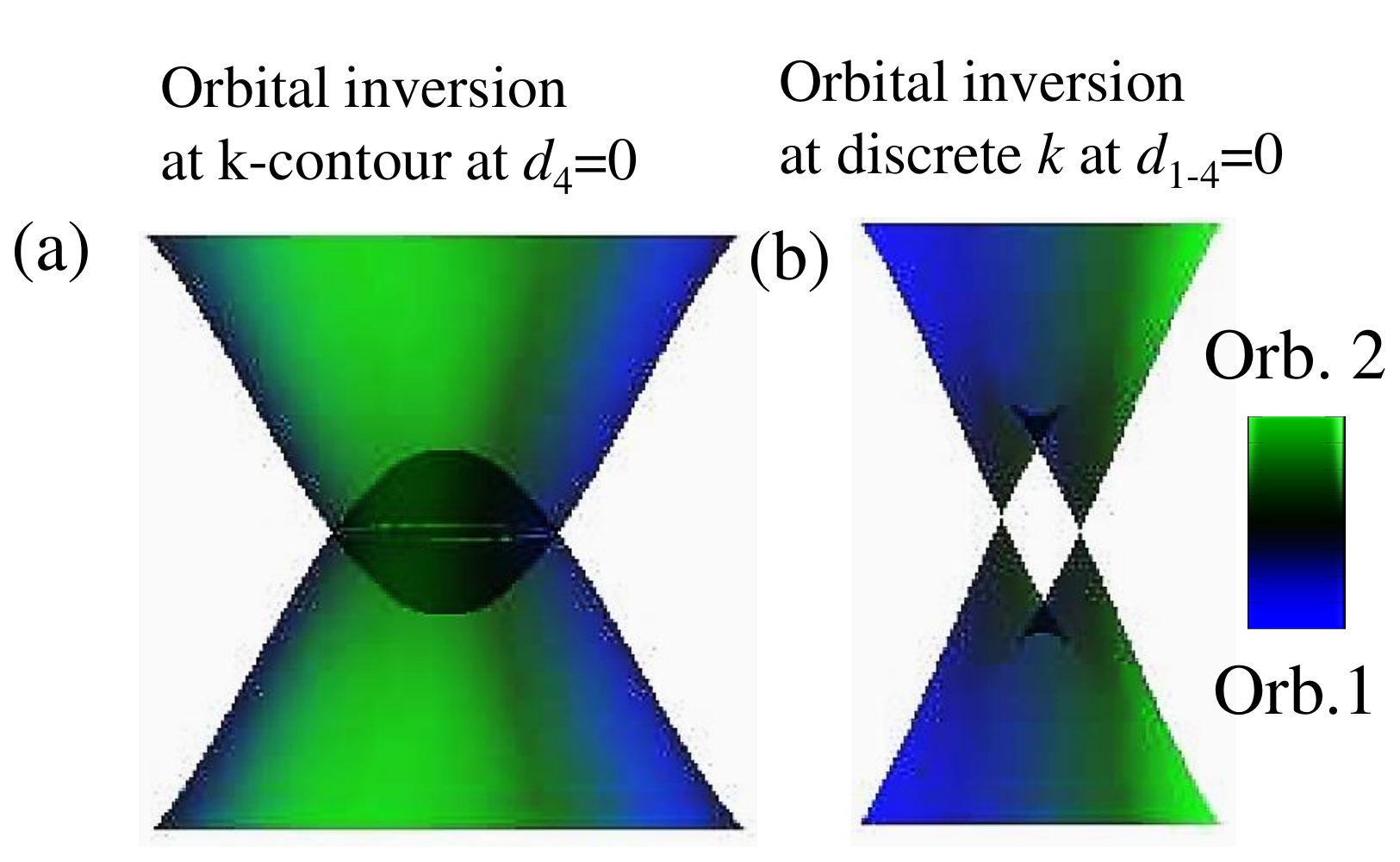}}
\caption{Schematic description of the formation of the Weyl orbital semimetal. (a) Contour of band touchings between two parabolic bands (without the inter-orbital hoppings). (b) As the odd-parity inter-orbital hoppings are turned on, Weyl nodes form at discrete momentum points.}
\label{fig0} 
\end{figure}

We derive the generalized theory of Weyl orbital semimetals starting from a multiband tight-binding model without invoking SOC: 
\begin{eqnarray}
H&=&-\sum_{s,n}\left[t_{s}^n c^{n\dag}_{s}c^n_{s\pm1}  + \mu^nc^{n\dag}_{s}c^n_{s}\right] \nonumber\\
&&-\sum_{s,\nu,n\ne m}  \nu \left [t^{nm}_{1s}c^{n\dag}_{s}c^m_{s+\nu1} + t^{nm}_{2s}c^{n\dag}_{s}c^m_{s+\nu2}\right].
\label{Eq1}
\end{eqnarray}
Here $c^{n\dag}_{s}~(c^{n}_{s})$ is the creation (annihilation) fermionic operator at a lattice site $s$, and $n,m$ are orbital indices. The first term is for intra-orbital hoppings which consists of only nearest neighbor hopping, and the second term is the chemical potential. For the inter-orbital hoppings, we consider both the nearest neighbor (third term) and the next-nearest neighbor hopping (fourth term). This is because, for various orbital and lattice geometries, either of them can be the leading term, and can give rise to linear dispersion. The symmetry of the Hamiltonian will remain unchanged when higher order hopping parameters are included as it will be evident later.

Since we are only interested in the inter-orbital hoppings which can give rise to linear dispersion, we restrict our discussion to the odd parity hopping between different orbitals. For this purpose, we have introduced an index $\nu=\pm$ which changes sign when the direction of hopping is reversed. This is the crucial part of our theory which can give imaginary hopping without SOC. With a Fourier transformation to the momentum space, we obtain 
\begin{equation}
H_{\bf k}=\sum_{n}\xi_n({\bf k}) c^{n\dag}_{\bf k}c^n_{\bf k} +\sum_{m\ne n} \xi_{nm}({\bf k}) c^{n\dag}_{\bf k}c^m_{\bf k}.
\label{eq:Hamk0}
\end{equation}
The first term in the above equation is the intra-orbital band dispersion for free fermions, while the second term consists of inter-orbital hopping integrals.  Without loosing generality, Eq.~\ref{eq:Hamk0} can be written in the basis of Dirac matrices $\Gamma_j$ as\cite{DasWeyl}
\begin{equation}\label{E:Hk}
H_{\bf k} =  \xi^+({\bf k})\Gamma_0+ \xi^-({\bf k})\Gamma_4 + \sum_{j=1}^3 d_j({\bf k})\Gamma_j,
\end{equation}
where $\xi^{\pm}({\bf k})=(\xi_1({\bf k})\pm\xi_2({\bf k}))/2$. $\xi_{{\bf k}}^-$ corresponds to $d_4$ components, and the remaining $d_j$ vectors consist of any combination of the inter-basis hopping $\xi_{nm}$. Given that the energy spectrum of $H_{\bf k}$ is $E^{\pm}({\bf k})=\xi^+({\bm k})\pm\sqrt{\sum_{j=1}^4 d_j^2({\bm k})}$, protected Dirac or Weyl nodes commence when all four $d_{1-4}$ components vanish simultaneously at discrete momenta ($\xi^+$ acts as the chemical potential shift to the overall band structure, and is not `in principle' required to vanish together). Around the nodal points, the above Hamiltonian can therefore be reduced to a general Weyl Hamiltonian\cite{Weyl} as $H_{\bf q}\psi_{\bf q}^{\pm} = \pm v_F {\bf q}.{\bf \Gamma}\psi_{\bf q}^{\pm}$ along three momentum directions, where $v_F$ is the Weyl Fermi velocity, ${\bf q}$ to be measured from the nodal point.  Here  $\psi^{\pm}$ corresponds to Weyl fermions possessing opposite chirality or winding numbers, and its net value vanishes in the whole Brillouin zone.\cite{WeylreviewBalents,Weylreview,Weylreview2} 

\begin{figure*}[top]
\rotatebox[origin=c]{0}{\includegraphics[width=1.95\columnwidth]{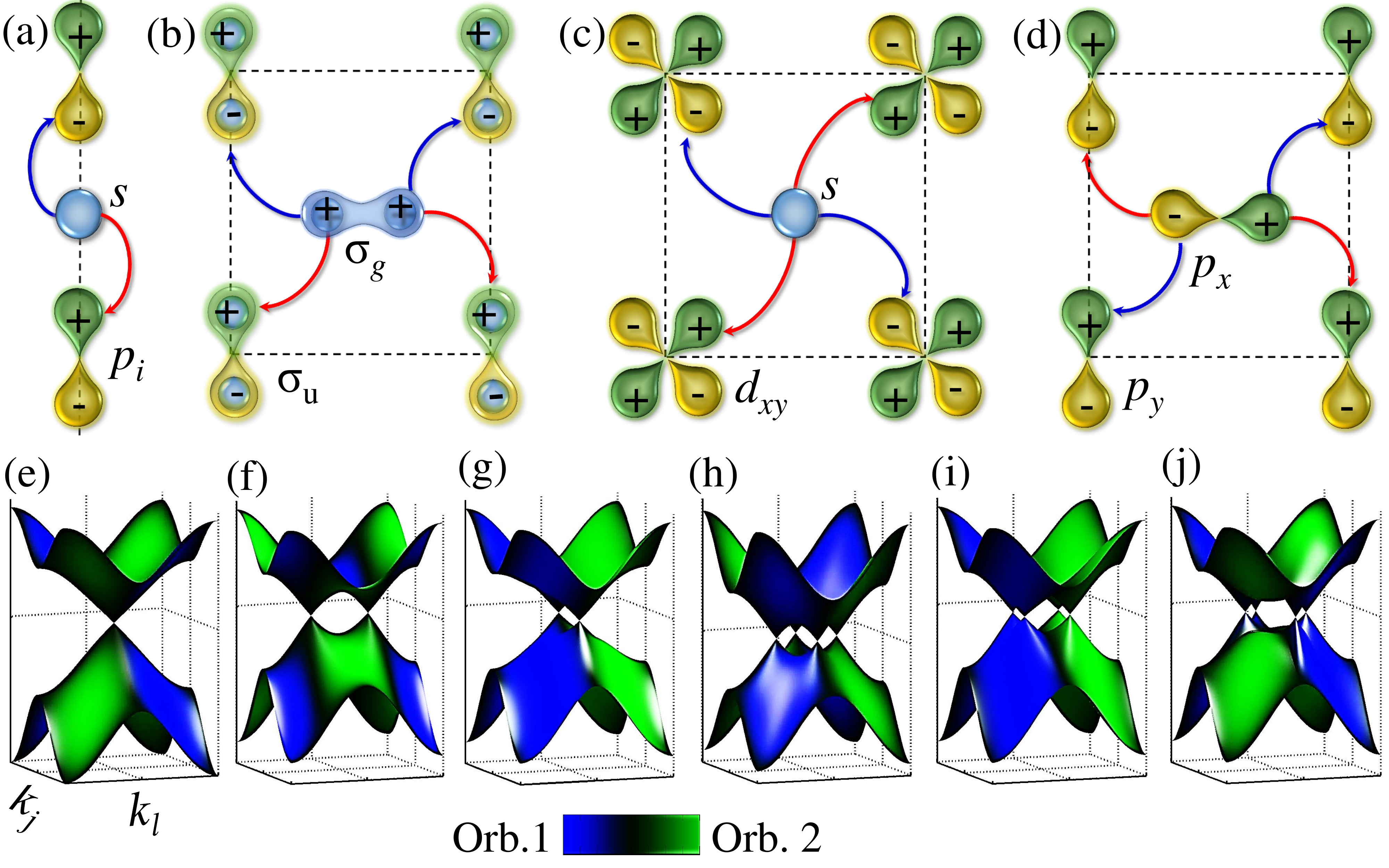}}
\caption{Various possible mechanisms and dispersion properties of Weyl orbital semimetals. (a, b) Even and odd orbitals or equivalently bonding and antibonding states lying along the zone boundary or diagonal directions, respectively, producing inter-orbital hoppings $\xi_{a/b}$. (c, d) For a pair of even or odd orbitals, a linear-in-momentum hopping term $\xi_c$ develops. The choices of orbitals for even and odd symmetries in this figure are representative and more such combinations can be easily thought of. Red and blue arrows depict same phase and out-of phase electron tunnelings, respectively. (e-j) Various illustrative cases of the formation of Dirac or Weyl nodes. In (e), both inversion and Mirror symmetry are held which produces a single 3D Dirac cone at the $\Gamma$-point. $\xi_{a/b}$ individually produce pairs of Weyl cones in (f, g), while $\xi_c$ produces four Weyl cones in (h, i, j). In (j), we find that Weyl cones are possible along the non-high-symmetric directions. $k_j$ and $k_l$ are any reciprocal lattice axes. Blue to green color scale depicts the orbital weight in the dispersion spectrum.}
\label{fig1} 
\end{figure*}

For brevity, we denote dispersion $\xi_j$ with numeric subscript as intra-orbital term ($\xi_{1,2}$), while English alphabetic subscripts give inter-orbital hopping terms ($\xi_{nm}=\xi_{a/b/c}$). For any pair of orbitals, the intra-orbital bands $\xi_{1,2}$ have parabolic dispersions, and the Weyl nodes can appear at the loci of $\xi_1=\xi_2$. This makes the Dirac mass ($d_4$) to vanish at a constant energy contour as shown in Fig.~\ref{fig0}(a), which occurs in materials due to band inversion, and thus it is parameter dependent. The remaining three $d$-vector components not only have to simultaneously vanish at discrete momenta on the same contour, but should also contain odd-parity inter-orbital hopping term and/or imaginary hopping term. Under such circumstances, the effective Hamiltonian near the nodal points can be expressed via Weyl Hamiltonian with linear dispersion, as demonstrated in Fig.~\ref{fig0}(b). When all $d_i$ vanish at high-symmetry $k$-points, the corresponding odd-number of cone is called 3D Dirac cone, and when cones form at non-high-symmetry $k$-points, they come in pairs with opposite orbital Chern numbers, which are called Weyl cones. 

The spinor basis for the Weyl fermions can be sought from multiple degrees of freedom, with one pair of different orbitals containing the above-mentioned inter-orbital hopping term, while the other two basis can arise from the sublattice symmetry, or from spin, or from the multiplets of the same orbitals. For example, in a $C_4$ symmetric lattice, $s$ and $p$ orbitals can form a Weyl orbital pair in which $p_i$ multiplets fill in the remaining basis, or between $p$ and $d$ orbitals in which $p_{x/y}$ and $d_{xz/yz}$ are all degenerate along the zone diagonal direction, and so on. Similarly,  owing to crystal inversion symmetry, a bonding state of two orbitals, and antibonding state of the same or different two orbitals can form a Weyl orbital pair. In Fig.~\ref{fig1}, we demonstrate several representative combinations of orbital and crystal symmetries for 3D Dirac or Weyl orbital nodes. Subsequently, four predicted materials exemplify different combinations of orbital symmetries which can lead to Weyl node formations. 
  
\subsection{Specific examples}

The intra-orbital band dispersions for the general case with nearest neighbor hopping become $\xi_{1,2}({\bf k}) = -2\sum_{j=x,y,z} t^{1,2}_j\cos{(k_ja_j)} - \mu^{1,2}$, and $a_j$ are the corresponding lattice constants. Here we discuss various different conditions under which the inter-orbital hopping terms $\xi_{a/b/c}$ can obtain linear dispersion purely from the angular dependence of the orbital symmetry, without any specific sublattice symmetry, or SOC. 

First we discuss a combination of even and odd symmetric states (under spatial inversion) placed either along the Brillouin zone boundary direction in Fig.~\ref{fig1}(a), or along the diagonal direction in Fig.~\ref{fig1}(b). Even and odd states can arise from $s$ or $d$, and $p$ or $f$ orbitals, respectively, or from bonding and antibonding combinations of any two orbitals or same orbitals from different sublattices (we use the case of the bonding and antibonding states for $s$ orbitals in the discussion because of its simplicity). In both cases, the odd parity orbitals give odd functional hoppings to the even state sitting at the center under reflection ($\nu=\pm$ in Eq.~\ref{Eq1}). The combination in Fig.~\ref{fig1}(a) guarantees the corresponding inter-orbital hopping matrix element to be $\xi_a(k_j)=2it_j\sin{(k_ja_j)}\sim 2it_ja_jk_j$ (near the nodal points). Here $i$ is the imaginary number, $j$ is the bonding direction, and $t_j$ is the hopping amplitude between the even and odd states separated by a distance of $a_j$, and $k_j$ is the lattice momentum.

Equivalently, when even and odd states are placed diagonally in a 2D $k_j-k_l$ plane (similar situation also arises in the 3D case) as shown in Fig.~\ref{fig1}(b), the resulting inter-band dispersion turns out to be linear along one momentum direction (which is odd under reflection, say $k_l$) as $\xi_b(k_j,k_l) = 4it_{jl}\cos{(k_ja_j)}\sin{(k_la_l)} \sim 4it_{jl}(1-\frac 1 2 a_j^2k_j^2)a_lk_l$. 

On the other hand, when we consider a pair of orbitals with same symmetry, their symmetry property prohibits the formation of linear dispersion along the zone boundary direction. However, when either two even or two odd orbitals are placed equidistantly along the diagonal directions, anisotropic and odd functional electronic hybridization may arise between them as demonstrated in Figs.~\ref{fig1}(c)-\ref{fig1}(d). For a pair of even orbitals, the linear hybridization can occur between an $s$ orbital and any of the $t_{2g}$ state of the $d$ orbital, or between the $t_{2g}$ and $e_g$ orbitals which are split in energy and momentum by either different occupation numbers, or crystal field splitting or other effects  as applicable. Since $t_{2g}$ orbitals intrinsically break crystal rotational symmetry, its orbital overlap term with an $e_g$ or $s$ state becomes an odd function of momentum. This is shown in Fig.~\ref{fig1}(c) for an example case of $d_{xy}$ and $s$ orbitals. Such a combination yields $\xi_c(k_j,k_l) = 4t^{\prime}_{jl}\sin{(k_ja_j)}\sin{(k_la_l)} \sim 4t^{\prime}_{jl}a_ja_lk_jk_l$. Interestingly, for the same multiorbital setup along the diagonal direction, two orthogonal odd orbitals, such as $p_x$, $p_y$ are allowed to hybridize. The corresponding dispersion, $\xi_c$, comes from a linear combination of $\pi$ and $\sigma$ bonds. It should be noted that $d_{xz}$ and $d_{yz}$ orbitals can mimic odd parity orbitals under inversion when projected onto the $x$--$y$ plane, or two antibonding states would also give a similar linear hopping term.

Different combinations of $\xi_{a,b,c}$ associated with different Dirac matrices $\Gamma_{1,2,3}$ determine the number and location of possible Dirac and Weyl cones. $\xi_a$ helps create nodal points along its propagating axis, while $\xi_b$ simultaneously produces massless and massive Dirac/Weyl terms along the $k_l$, and $k_j$ directions, respectively. The resulting Weyl nodes always come in multiple of two or merge into a 3D Dirac cone at the $\Gamma$ point. On the other hand, $\xi_c$ term produces Weyl nodes in multiple of $n$, in systems with $C_n$ rotational symmetry, and maintain the translational symmetry of the lattice. Some other combinations of $\xi_{a,b,c}$, however, can sometimes gap out each other's nodal states. 

Some examples of such large possibility of Weyl orbital nodes in various setup are shown in Fig.~\ref{fig1} (lower panel). In Fig.~\ref{fig1}(e), we present a single 3D Dirac cone at the $\Gamma$ point for an illustrative combination of $d_j=-i\xi_a (k_j)$ ($j$ = $x$, $y$, $z$). The 3D Dirac cone splits into pairs of Weyl points as the inversion symmetry is lifted. Pair of Weyl cones can be created by using either both $\xi_{a/b}$ terms (Fig.~\ref{fig1}(f)), or one of them and combine it with $\xi_c$ (Fig.~\ref{fig1}(g)) which results in contrasting orbital texture inversions. Four Weyl cones can be created with various combinations; for example, for a combination of $\xi_a$ and $\xi_c$, and so on, as shown in Fig.~\ref{fig1}(h). These four Weyl nodes can be rotated from the bonding directions toward the zone diagonal ones by breaking the corresponding symmetry assigned with the $\Gamma_3$ matrix. Using the same combination as in Fig.~\ref{fig1}(e), we get four Weyl cones along the zone diagonal directions when $d_1= \xi_a(k_n)$, $d_2=\xi_a(k_j)$, and  $d_3=\pm i\xi_a(k_l)$ ($k_n$ is perpendicular direction to the $k_j,~k_l$ plane), see Fig.~\ref{fig1}(i). Finally, Weyl cones can appear along non-high-symmetry directions,  Fig.~\ref{fig1}(j), for a choice of $d_1+id_2=-(i/2)\xi_c(k_j,k_l)$ and $d_3=\pm i[\xi_a(k_j)-\xi_a(k_l)]$. However, such Weyl nodes are not protected and can be gapped by disorder or perturbations. The details of the parameters used in the above presentation are listed in the Appendix.

The above examples demonstrate how two entangled degrees of freedom arise purely from the orbital texture inversion at discrete momentum points for spinless fermions. Due to the conservation of orbital angular momentum across the Weyl nodes, they remain time-reversal invariant. In these senses, our theory of Weyl orbital semimetal is different from the hexagonal symmetry related graphene,\cite{CastroNeto} or one-dimensional polyacetylene,\cite{SSH_model} or from mirror symmetry induced topological crystalline insulator.\cite{TCI} Opening a band gap at the Weyl orbital points can lead to Weyl orbital topological insulators, which is reminiscence of weak $Z_2$ topological insulator in time-reversal invariant systems.\cite{Hasanreview,Zhangreview}  

\section{{\it Ab-initio} calculations and materials predictions}
Electronic structure calculations are carried out by DFT method with the generalized gradient approximation (GGA) in the parametrization of Perdew, Burke and Ernzerhof (PBE)~\cite{PBE} as implemented in the Vienna {\it ab-initio} simulation package (VASP)~\cite{VASP}. Projected augmented-wave (PAW)~\cite{PAW} pseudo-potentials are used to describe core electrons. The conjugate gradient method is used to obtain relaxed geometries. We include Hubbard $U$ corrections to the GGA calculations (GGA+$U$) and consider the spherically averaged form of the rotationally invariant effective $U$ parameter with $U$ = 3.0 eV, 5.0 eV and 4.0 eV on the correlated V 3$d$, Ni 3$d$ and Ti 3$d$ orbitals, respectively. In the cases where GGA+ $U$ is not considered, i.e. for BLi and PbO$_2$, the calculation is repeated with a different Heyd-Scuseria-Ernzerhof (HSE06)~\cite{HSE06} hybrid functional to check for the band crossings, and found that the gapless Weyl cones are intact. In the HSE06 calculations, the same geometry as the PBE cases are used. For each system, we have explicitly checked that the SOC does not gap out the bulk Weyl nodes. Both atomic positions and cell parameters have been allowed to relax, until the forces on each atom are less than 5 meV/\AA. The optimized lattice vectors and internal coordinates of all the atoms are listed in supplementary Table~SI. We chose both the electronic wavefunction-cutoff energy and $k$-mesh (for Brillouin zone sampling) such that the accuracy of a total energy convergence is less than 10$^{-4}$ eV/unit-cell.

The structural stabilities of these materials are investigated by calculating the phonon dispersion and the cohesive energy, $E_{\mathrm {coh}}$. Force constants are calculated for a 2$\times$2$\times$2 super-cell within the framework density functional perturbation theory~\cite{DFTP} using the VASP code. Subsequently, phonon dispersions are calculated using phonopy package~\cite{phonopy}, and the results are shown in supplementary Fig.~S1. Finally, we find that all materials have negative cohesive energy, implying that structure can exist in bound state.

\subsection{3D `graphene' from sublattice symmetry}

\begin{figure*}[ht]
\rotatebox[origin=c]{0}{\includegraphics[width=1.9\columnwidth]{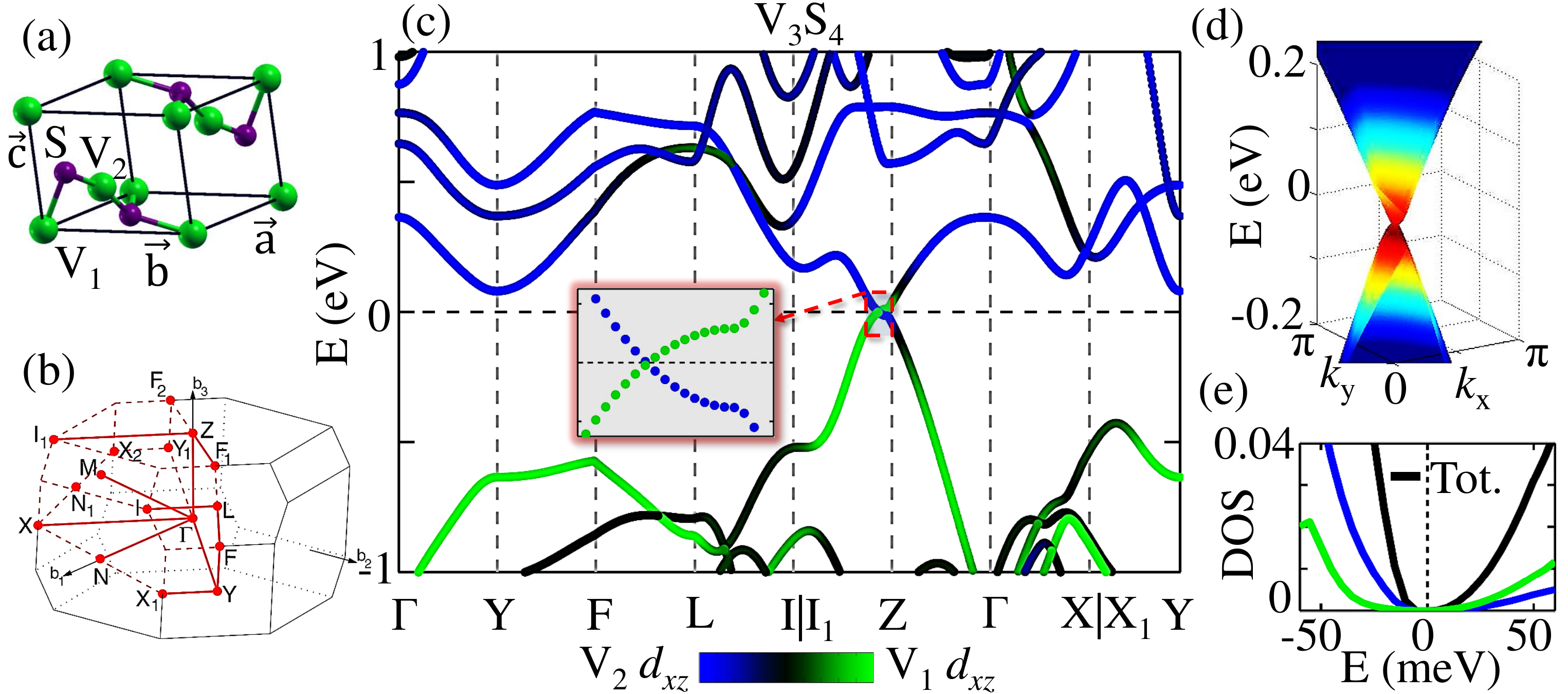}}
\caption{Band structure and DOS of V$_3$S$_4$. (a) The crystal structure of V$_3$S$_4$, and (b) the corresponding reciprocal space. (c) Electronic band structure is shown along the high-symmetry momentum directions, with a blue to green color gradient map which dictates the corresponding orbital weight from two inequivalent V atoms. (d), Full dispersion of the two low-lying bands on the `Z' plane, exhibit a single Weyl point along Z-I$_1$ direction in the first quadrant. The blue to red color map has not meaning in this figure. (e), Low-energy DOS (black color), and partial DOS (blue and green) for the same two orbital states. The DOS is given in the unit of number of states per unit eV.}\label{figV3S4}
\end{figure*}

We first discuss the band structure of the existing V$_3$S$_4$ compound,\cite{V3S4_Kawada,V3S4_Mujica} and the origin of bulk Weyl cones in this material. V$_3$S$_4$ has a monoclinic phase in the space group of $C2/m$ (No. 12), Fig.~\ref{figV3S4}(a). A susceptibility measurement in the polycrystalline sample of V$_3$S$_4$ reported an antiferromagnetic (AFM) transition with N\'eel temperature $T_N\sim$ 9~K and magnetic moment of $\sim$ 0.2~$\mu_{\mathrm B}$ per V atom, and also a ferromagnetic (FM) phase below $T\sim$ 4.2~K.\cite{V3S4_AFM}. Our GGA+$U$ calculation shows that the FM phase has the lowest ground state energy (by 36 meV/f.u. with a magnetic moment of 2.1~$\mu_{\mathrm B}$ per V atom). In the FM phase, our band structure calculation in Fig.~\ref{figV3S4}(c) shows that all other $k_z$-planes are 2D band insulators, except the $k_z=\pi/c$ one, which has a well-defined Weyl cone for the same spin state. Two inequivalent vanadium atoms are placed in the corner and interior basis of the lattice, as shown in Fig.~\ref{figV3S4}(a), which obtain equal electron occupation number. Therefore, their same $d$-orbitals osculate near the Fermi level, which promote their inter atomic hybridization to follow the even-even orbital hopping principle described in Fig.~\ref{fig1}(c) above. Interestingly, the characteristic band dispersion of V$_3$S$_4$ resembles 2D graphene,\cite{CastroNeto} despite the differences in the crystal structure and orbital contributions between them. Another characteristic difference between the 2D Dirac cone and bulk Weyl node is that in the former case, the corresponding density of states (DOS) is linear-in-energy across the Dirac cone, while here it is quadratic in energy.\cite{Weylreview} This is indeed evident  in the computed DOS of V$_3$S$_4$, presented in Fig.~\ref{figV3S4}(e), overlayed with the partial DOSs to demonstrate how the DOSs of the two V atoms change across the Weyl node.
%

\subsection{Weyl node induced by two even orbitals}
\begin{figure*}[ht]
\rotatebox[origin=c]{0}{\includegraphics[width=1.9\columnwidth]{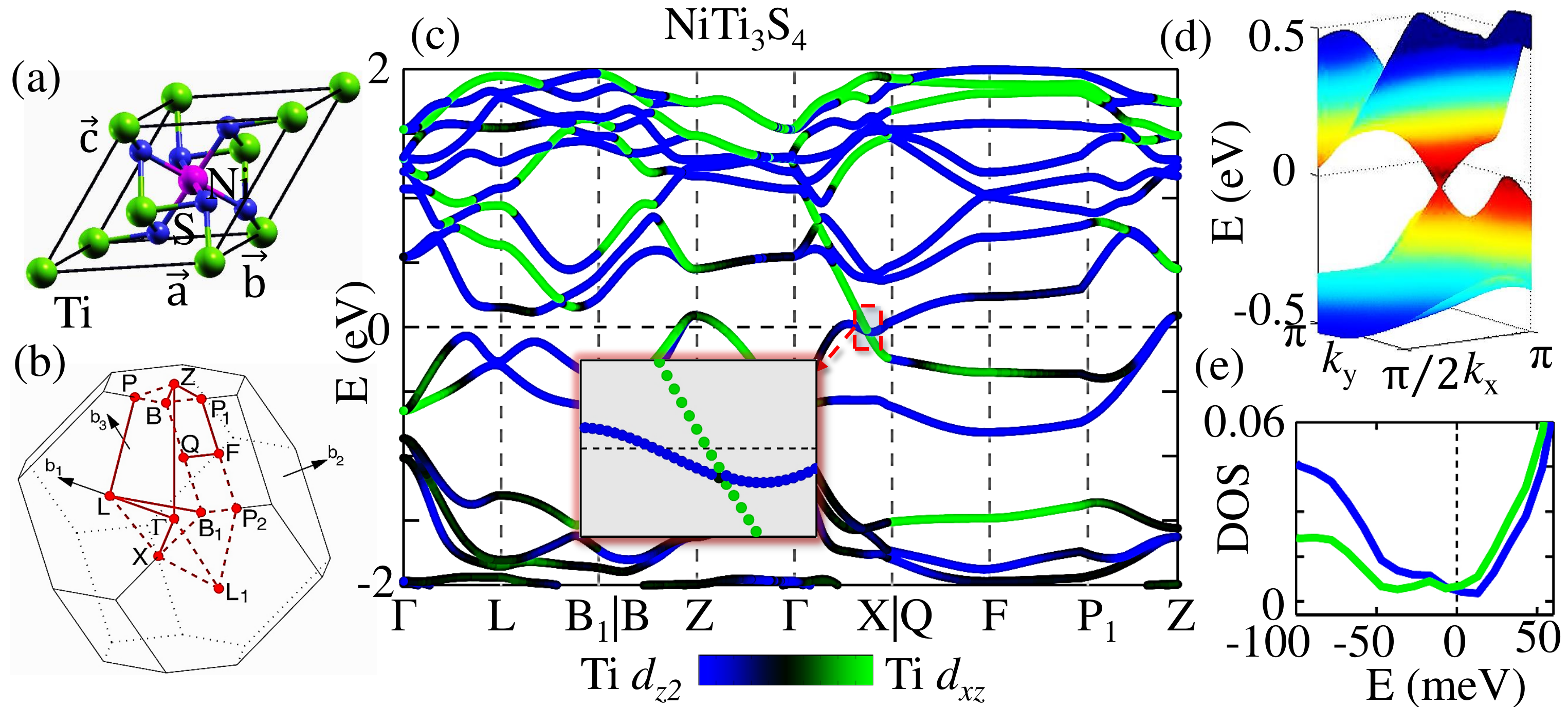}}
\caption{Band structure and DOS of NiTi$_3$S$_6$. (a)-(b), Real space and reciprocal space representations of the crystal structure. (c), (d), (e), Corresponding electronic band structure along high-symmetry directions, in 2D plane for the two lowest energy bands, and low-energy DOS, respectively. }
\label{figNiS2}
\end{figure*}
Our next example is a ternary transition metal sulphide NiTi$_3$S$_6$, in which a unusual anisotropic crystal field splitting of the Ti $d$-orbitals vanishes at discrete momenta and forms Weyl nodes, see Fig.~\ref{figNiS2}. NiTi$_3$S$_6$ is a known material with a lattice structure belonging to the rhombohedral lattice with space group of R-3H (No. 148).\cite{NiTi3S6} Ti has two electrons in its 3$d$ orbitals, which are shared between the conduction $t_{2g}$ and valence $e_{g}$ orbitals, separated by the crystal field splitting of the rhombohedral lattice. To delineate the origin of odd functional dispersion from the overlap matrix-element between the two even orbitals, let us focus on the $a-b$ plane of the lattice. The projected $d_{xz}$ orbital on the $x$--$y$ plane mimics the $p_x$-type orbital in the sense that its phase changes sign on both sides of the center position. On the other hand, $d_{z^2}$ orbital acts as a purely isotropic orbital on this plane. Therefore, the inter-orbital electron tunneling between them follows the principle depicted in Fig.~\ref{fig1}(a), and creates orbitally polarized Weyl nodes. Our predicted Weyl cones in this material may perhaps be less compelling for functional use, however, it exemplifies a new mechanism of the Weyl orbital semimetal arising from the least expected momentum dependent crystal field splitting. However, strain or pressure can be applied to remove the additional electron/hole pockets from the Fermi level, if needed.

\subsection{Weyl node induced by two odd orbitals}

\begin{figure*}[ht]
\rotatebox[origin=c]{0}{\includegraphics[width=1.9\columnwidth]{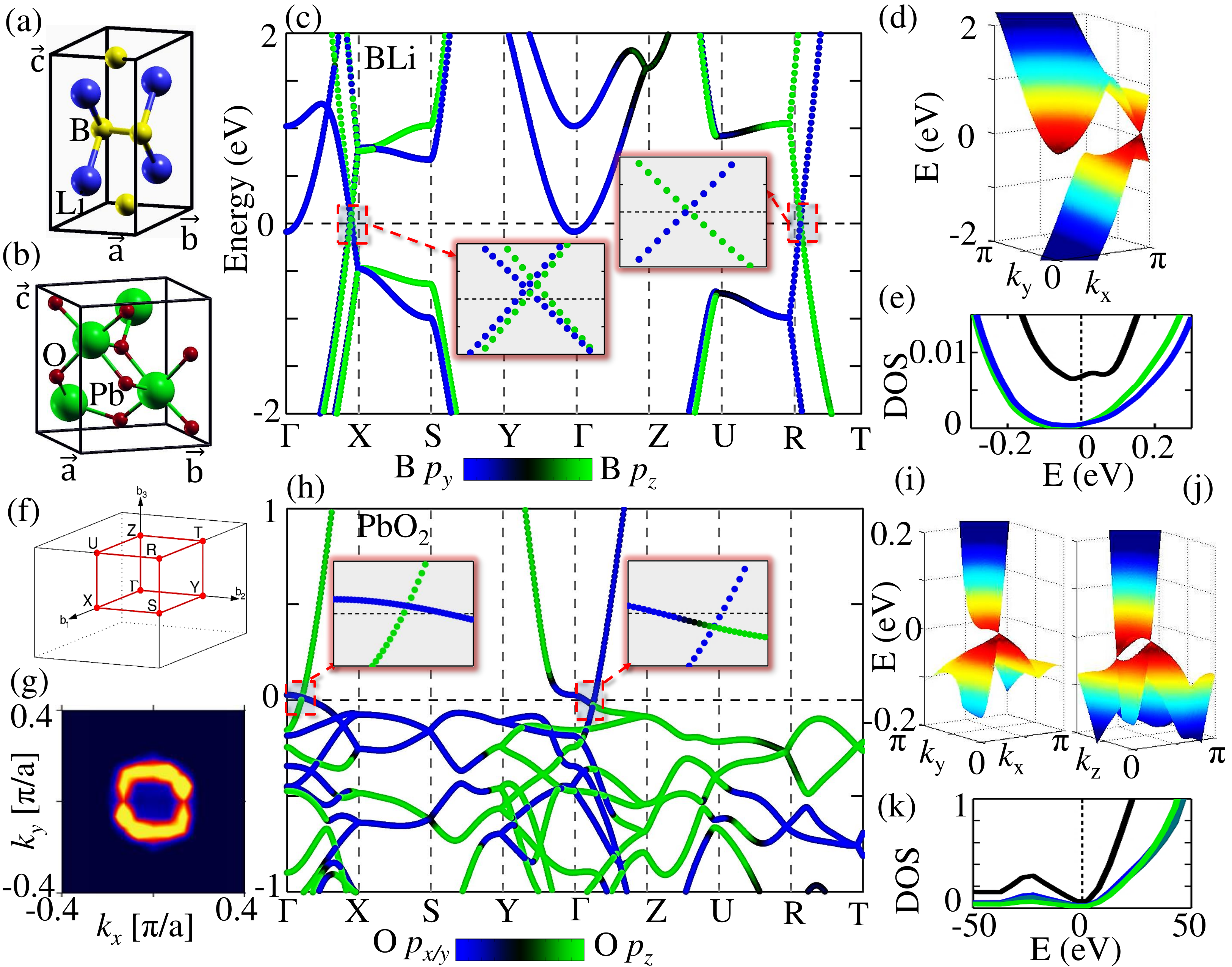}}
\caption{Band structures and DOSs of isostructural BLi and PbO$_2$. Upper panel shows results of BLi, while the bottom panel gives the same results for PbO$_2$. In both systems, the orbital characters interchange between the $p_{x/y}$ and $p_z$ orbitals of Li and O atoms, respectively, via protected Weyl points. Large dispersions of these light elements give rise to vanishingly small DOS, which is almost zero above the Fermi level in PbO$_2$ in (e). (g) Spectral weight map of the surface state of PbO$_2$ in (001) cut exhibits `Fermi arc' behavior.
}\label{figBLiPbO2}
\end{figure*}

Finally, we discuss two materials BLi and PbO$_2$ in Fig.~\ref{figBLiPbO2} which show multiple Weyl cones, forming at the crossing points of two $p$ orbitals states in their native phases. Both materials form orthorhombic lattice.\cite{PbO2} Owing to the orthorhombic structure, the orthogonality between the $p_{x/y}$ and $p_z$ orbitals is lifted and the inter-orbital electronic interaction is turned on. The bonding between the two orthogonal orbitals stems from a linear combination of both $\pi$ and $\sigma$ bondings, and thus is generally robust against strain. Our band structure calculations show that multiple Weyl cones are formed along the Brillouin zone boundary at $k_z=0$ and $k_z=\pi$ planes. The projected $p_z$ states on the $x$--$y$ plane appears isotropic to the $p_{x/y}$, which therefore promotes the corresponding inter-orbital hopping to follow the scenario described in Fig.~\ref{fig1}(a) and/ or Fig.~\ref{fig1}(b). 

In BLi, there are two Weyl cones almost overlapping with each other along the $\Gamma$-X direction, which are thus less stable to impurity scattering. In PbO$_2$, Weyl  cones arise from O atoms and thus remain ungapped even after including the SOC. Because of the small atomic number of B and O atoms, the corresponding Weyl states are highly dispersive, and the corresponding DOSs are reduced, see Fig.~\ref{figBLiPbO2}(e) and \ref{figBLiPbO2}(k). We find that the extra electron/hole-pockets from the Fermi level in BLi are removed using HSE06 functional (not shown). Furthermore, the band inversion strength for PbO$_2$ is relatively weak. Using relaxed lattice constants, we find Weyl nodes in the present GGA calculation, while HSE06 functional gives a 3D Dirac cone, and the MBJ functional looses the band inversion. However, with small strain of about 3\%, we find that the latter two functionals also give Weyl nodes along the same momentum directions as in GGA band structure.

The bulk-boundary correspondence associated with the general form of Weyl Hamiltonian in transnational invariant lattice dictates that the Fermi surface on the edge state becomes disconnected at the points where the bulk and edge states merge, creating truncated Fermi surface (s) or the so-called `Fermi arc'.\cite{Weylreview,Weylreview2} To ascertain that our predicted materials indeed belong to the Weyl semimetal class, we have computed the surface state for a representative material PbO$_2$ because of its structural simplicity. The slab of PbO$_2$ are modeled with 001 surface containing 16 atomic layers of Pb. A 14~\AA  vacuum is place at the surface to avoid the interaction between two consecutive supercell. The spectral weight map of the Fermi surface of the corresponding surface state indeed shows a `Fermi arc' in Fig.~\ref{figBLiPbO2}(g).

\section{Weyl Fermi velocity and possibility of orbitronics}
\begin{figure}[top]
\centering
\rotatebox[origin=c]{0}{\includegraphics[width=0.90\columnwidth]{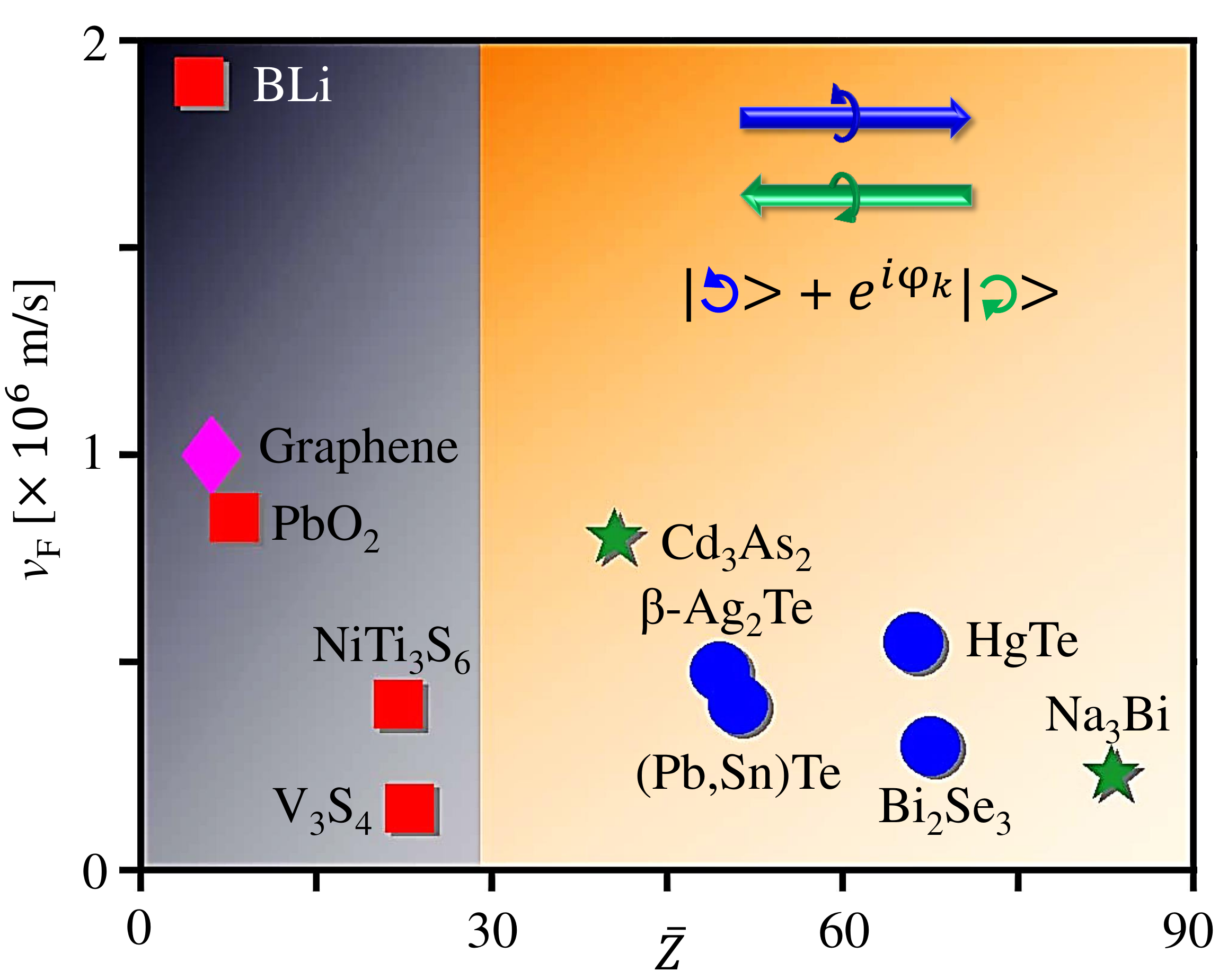}}
\caption{Fermi velocity of various classes of Dirac materials. Computed Fermi velocity at the Dirac cone (averaged over the two intersecting linear-dispersion) of the four Weyl orbital semimetals predicted here are compared with various other experimentally verified Dirac materials. All SOC induced Dirac fermions in heavy-elements have Fermi velocity almost an order of magnitude lower than that of the Weyl orbital semimetals, and graphene. The horizontal coordinate gives the average atomic number ($\bar{Z}$) of the elements contributing to the Dirac cone. Gray and yellow shadings separate the two families of Dirac materials without and with SOC, respectively. The Fermi velocity data are taken for the surface states of the 2D topological insulator HgTe/CdTe from Ref.~\cite{HgTe}, for the 3D topological insulator Bi$_2$Se$_3$ from \cite{Bi2Se3}, and  for the topological crystalline insulators (Pb,Sn)Te from Ref.~\onlinecite{SnTeAndo,PbSnTe}, $\beta$-Ag$_2$Te from Ref.~\onlinecite{Ag2Te}.  The Fermi velocity at the 3D Dirac cone of the Weyl semiletals Cd$_3$As$_2$ is taken from Ref.~\onlinecite{ExpCdAsHasan,ExpCdAsBorisenko}, and for  Na$_3$Bi from Ref.~\onlinecite{ExpNa3Bi,ExpNa3BiHasan}. The data for the non-SOC induced Dirac cone in graphene is taken from Ref.~\onlinecite{CastroNeto}. The inset figure schematically shows the possibility of obtaining orbitally polarized electronic current with an anisotropic phase difference, $\phi_k$, protecting their quantized currents.
}\label{figvF}
\end{figure}

In Fig.~\ref{figvF}, we compare the velocity of all four Weyl orbital semimetals predicted here, with the other known Dirac or Weyl materials. All Dirac materials having Dirac cones generated as a manifestation of the SOC consist of `post-transition metals' with higher atomic numbers. With increasing atomic number, the bandwidth decreases and the effects of correlation increases, which all conspire in a gradual reduction of the Fermi velocity. On the other hand, the Weyl orbital semimetals are applicable to any combinations of orbitals, and does not depend on the lattice or atomic properties such as bulk band gap or SOC. This flexibility greatly helps expanding the territory of Dirac materials to very light atoms such as Li, B, C, O, S, V, Ti and Ni with Weyl fermion velocity larger by an order of magnitude. In fact, the Fermi velocity for BLi is found to be $\sim$2$\times$10$^{6}$m/s, which is the highest among all Dirac materials known to date.

Any two-component quantum degrees of freedom of electron in a lattice can mathematically behave in the same way as spin does, and therefore they can be viewed as pseudospin. In this sense, two interlocked orbital degrees of freedom with texture inversion enter into the low-energy Dirac equation in precisely the same manner that real spin produces Dirac equation in topological insulators, or Weyl semimetals, or the pseudospin behaves in the graphene's  Dirac equation. Therefore, the physical concepts relating to the spintronics, spin-orbitronics  or pseudospintronics application also apply to the Weyl orbital semimetals. We expect orbitally polarized charge current or orbital current in the bulk of these materials, which are protected by a momentum-dependent phase difference $\phi_k$ (see inset to Fig.~\ref{figvF}), generic to any Dirac electron motions.\cite{CastroNeto,Hasanreview,Zhangreview,pseudospintronics} The corresponding Berry curvatures for the two orbitals pick up opposite values, which can lead to quantum orbital Hall effect. The nature of impurity scattering protection for different quantum orbitronics cases is characteristically different. In the present family, an electron can only scatter from one orbital state to another when the impurity vertex contains a corresponding anisotropic orbital-exchange matrix-element or if the electron dynamically passes through the momentum and energy of the Dirac cone. Another advantage of the Weyl orbital semimetal is that here the Dirac cone is even immune to time-reversal symmetry breaking, and a bulk gap can be engineered by the lattice distortion. Therefore, the generation, transport and detection of orbitally protected electric current may lead to new opportunities for orbitronics. Chiral orbital current in the Weyl semimetals can be detected by Kerr effect. Possibilities of orbitronics were predicted earlier in $p$-doped silicon\cite{orbitronicsSi} and $p$-doped graphene,\cite{orbitronicsgraphene} which are yet to be observed. 

\section{Conclusions} 
The presented new family of Weyl orbital semimetal has the ability to bypass many limitations imposed in other Dirac materials including atomically thin graphene, topological insulators and Weyl semimetals under time-reversal invariance and SOC. An important advantage of the 3D Dirac cone than its 2D counterpart is that in the former setup the DOS is quadratic in energy which will be beneficial for designing faster transistors and hard drive with low energy consumption. The accessibility of the Dirac fermions in topological insulators is subjected to the bulk band gap, which can be filled by thermal broadening. Such a limitation is not present in Weyl semimetal families.\cite{WeylreviewBalents,Weylreview} Due to the absence of these constraints, the possible materials classes for the Weyl orbital states are, in principle, expanded to the entire periodic table. Moreover, both mechanical and chemical tuning induced band gap in the Weyl node are appealing features which can be useful to exciton condensation, photovoltaics and solar cell applications, and optoelectronic technology.\cite{Weylexciton} Finally, the discovery of Weyl orbital semimetals will open the door for a new field of orbitronics.

\vskip0.1cm
\noindent\\
{\bf Acknowledgments:} We thank Hsin Lin, Su Ying Quek, Matthias Graf for valuable discussions and hospitality. The work is facilitated by NERSC computing allocation in the USA and the the GRC high-performance computing facility in Singapore.

\appendix
\section{Parameter sets for Fig. 2}

We use Dirac matrices of the form $\Gamma_{1,2,3}=\sigma_1\otimes\sigma_{1,2,3}$, and $\Gamma_4=\mathcal{I}\otimes\sigma_3$, where $\sigma_i$ are the Pauli matrices and $\mathcal{I}$ is 2$\times$2 unity matrix. 

For the demonstration of the emergence of Dirac or Weyl ferminons, we take a simple and minimal set of parameters for $t^n$, $\mu^n$, and $t^{nm}$: $t_j^{n=1,2}=\pm$150~meV, and $t^{n\ne m}_{jl}$ = 150~meV is taken to be same for all orbitals $n$, $m$ and along any directions $j$, $l$. The chemical potential can be chosen in a way that $\xi_{\bf k}^-$ banishes at the $\Gamma$ point ($\mu^n=-6t^n$) or at any other discrete momenta ($\mu^n=-6t^n\pm\delta$, where $\delta$ is a tunable number). In Fig.~1 of main text, we take $\mu^{1,2}=\mp$0.9~eV for the Dirac point at the $\Gamma$, and $\mu^{1,2}=\mp$0.7~eV otherwise. All tight-binding parameters are kept same for all plots in Fig.~1.

We explicitly write down the combinations of $\xi_{a,b,c}$ chosen in Fig.~1 of the main text. In the following cases, we assume Dirac or Weyl cones are present in the $k_j$ and $k_l$ plane, and $k_n$ is the perpendicular axis. For Fig.~1E, the $d$-vectors are taken to be $d_{j} = -i\xi_a(k_j)$, where $j=1,2,3$ corresponding to $k_j$, $k_l$ and $k_n$ direction, or their various combinations.  The choice of ${\bf d}$-vector components are 
\begin{eqnarray}
&&{\rm For~Fig.~1(f):}~~~ d_1+id_2 = \frac{1}{2}\xi_b(k_j,k_l),~d_3 = \frac{1}{2}\xi_c(k_j,k_l), \nonumber\\
&&~~~~{\rm or}~ d_3 = -i\xi_a(k_n),~{\rm o    r}~ d_3 = -\frac{i}{2}\left[\xi_b(k_j,k_l)+\xi_b(k_n,k_l) \right].\nonumber\\
&&{\rm For~Fig.~1(g):}~~~ d_1+id_2 = \xi_a(k_l),~ d_3 = \frac{1}{2}\xi_c(k_j,k_n).\nonumber\\
&&{\rm For~Fig.~1(h):}~~~ d_1+id_2=\xi_a(k_n), ~ d_3 = \frac{1}{2}\xi_c(k_j,k_l),\nonumber\\
&&~~~~~~~~~~~~~~~~{\rm or}~d_1+id_2 = \frac{i}{2}\xi_c(k_j,k_l), ~ d_3=-i\xi_a(k_n).
\end{eqnarray}
The above three cases give Weyl cones along the zone axis. We also provide two other cases, where Weyl cones appear along other directions when a point-group symmetry is broken.  In these cases, both inter-basis hoppings between 1 to 3 and 2, 3 are taken to have same sign, violating the symmetry associated with the $\Gamma_3$ term. Such Weyl cones are probably not as stable as others. 
\begin{eqnarray}
&&{\rm For~Fig.~1(i):}~~~ d_1+id_2 = \left[\xi_a(k_n)+i\xi_a(k_j)\right],\nonumber\\
&&~~~~~~~~~~~~~~~~~~~~~~~ d_3 = \pm i\xi_a(k_l).\nonumber\\
&&{\rm For~Fig.~1(j):}~~~ d_1+id_2 = -\frac{i}{2}\xi_c(k_j,k_l),\nonumber\\
&&~~~~~~~~~~~~~~~~~~~~~~~~ d_3 = \pm i\left[\xi_a(k_j)-\xi_a(k_l)\right].
\end{eqnarray}

\begin{figure}[here]
\centering
\rotatebox[origin=c]{0}{\includegraphics[width=.95\columnwidth]{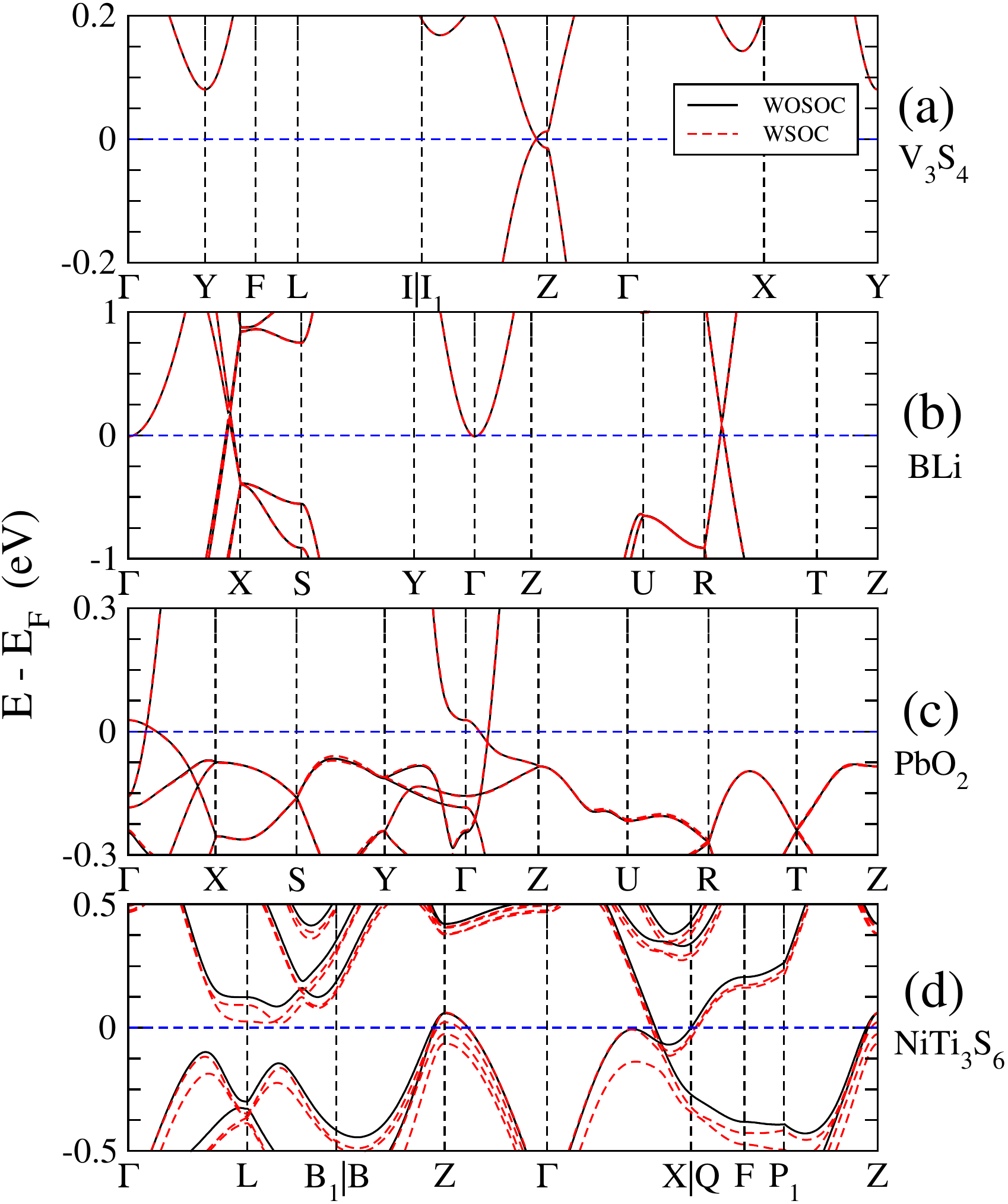}}
\caption{{\bf Supplementary Figure S1: Comparison of band structure with and without spin-orbit coupling}. Here we show the same band structure, plotted in the main figure, for all four materials studied in this paper, but with and without including the spin-orbit coupling effect. Expectedly, spin-orbit coupling in these materials is small enough to gap the orbital degenerate point at the Weyl nodes, which provide further justification that the Weyl nodes are protected. 
}\label{figbandwsoc}
\end{figure}

%
\begin{figure}[here]
\centering
\rotatebox[origin=c]{0}{\includegraphics[width=.99\columnwidth]{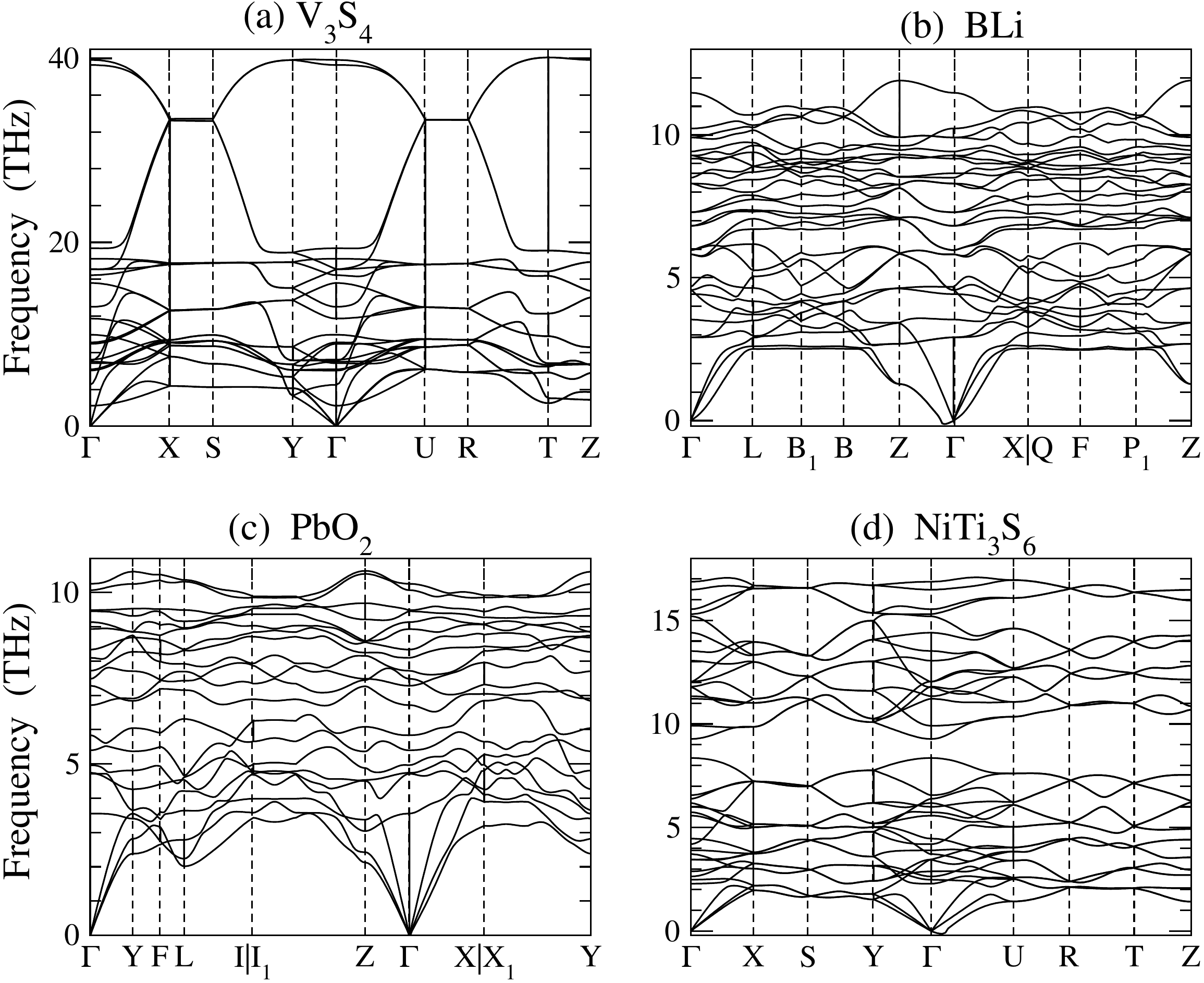}}
\caption{{\bf Supplementary Figure S2: Phonon spectrum}. Since some of the materials we predict here are yet to be synthesized experimentally, we carry out the phonon spectrum calculation to study the stability of the lattice. For all materials, we find a stable phonon spectrum.
}\label{figphonon}
\end{figure}


\begin{widetext}
\begin{table*}[h]
\centering
\begin{tabular}{p{2cm}| p{7cm} | c c c c}
\hline \hline 
Material & Crystal structure & Atoms & x & y & z \\
\hline
           & $a$=6.65987~\AA, $b$=6.65987~\AA, $c$ = 6.19743~\AA; & V1 & 0.000 & 0.000 & 0.000 \\  
           & $\alpha$ = 64.050$^{\circ}$, $\beta$ = 64.050$^{\circ}$, $\gamma$ = 30.759$^{\circ}$ & V2 & 0.737 & 0.737 & 0.285 \\ 
           & & V3 & 0.263 & 0.263 & 0.715 \\
V$_3$S$_4$ &  & S4 & 0.637 & 0.637 & 0.019 \\ 
           & & S5 & 0.363 & 0.363 & 0.981 \\ 
           & & S6 & 0.120 & 0.120 & 0.551 \\ 
           & & S7 & 0.880 & 0.880 & 0.449 \\ 
\hline
           & $a$= 6.68709~\AA, $b$= 6.68709~\AA, $c$ = 6.68709~\AA; &  Ni1 & 0.500 & 0.500 & 0.500 \\  
           & $\alpha$ = 52.977$^{\circ}$, $\beta$ = 52.977$^{\circ}$, $\gamma$ = 52.977$^{\circ}$ &  Ti2 & 0.000 & 0.000 & 0.000 \\ 
             &  &  Ti3 & 0.671 & 0.671 & 0.671 \\ 
             &  &  Ti4 & 0.329 & 0.329 & 0.329 \\ 
NiT$_3$S$_6$ &  &   S5 & 0.093 & 0.415 & 0.747 \\  
             &  &   S6 & 0.415 & 0.747 & 0.093 \\  
             &  &   S7 & 0.747 & 0.093 & 0.415 \\  
             &  &   S8 & 0.907 & 0.585 & 0.253 \\  
             &  &   S9 & 0.585 & 0.253 & 0.907 \\  
             &  &  S10 & 0.253 & 0.907 & 0.585 \\
\hline
          & $a$= 5.11933~\AA, $b$ = 5.57092~\AA, $c$ = 6.05971~\AA; &   Pb1  &  0.000 & 0.250 & 0.177 \\ 
          & $\alpha$ = 90.000$^{\circ}$, $\beta$ = 90.000$^{\circ}$, $\gamma$ = 90.000$^{\circ}$ &   Pb2  &  0.500 & 0.750 & 0.323 \\  
          &  &   Pb3  &  0.000 & 0.750 & 0.823 \\ 
          &  &   Pb4  &  0.500 & 0.250 & 0.677 \\ 
          &  &   O5   &  0.734 & 0.429 & 0.405 \\  
PbO$_2$   &  &   O6   &  0.766 & 0.929 & 0.095 \\  
          &  &   O7   &  0.266 & 0.071 & 0.405 \\  
          &  &   O8   &  0.234 & 0.571 & 0.095 \\  
          &  &   O9   &  0.266 & 0.571 & 0.595 \\  
          &  &   O10  &  0.234 & 0.071 & 0.905 \\ 
          &  &   O11  &  0.734 & 0.929 & 0.595 \\ 
          &  &   O12  &  0.766 & 0.429 & 0.905 \\ 
\hline
        & $a$ = 3.07497~\AA, $b$ = 5.67574~\AA, $c$ = 6.13950~\AA; &   B1 & 0.750 & 0.473 & 0.500\\  
        & $\alpha$ = 90.000$^{\circ}$, $\beta$ = 90.000$^{\circ}$, $\gamma$ = 90.000$^{\circ}$ &   B2 & 0.250 & 0.973 & 0.000\\  
        & &   B3 & 0.250 & 0.527 & 0.500\\  
BLi     & &   B4 & 0.750 & 0.027 & 1.000\\  
        & &  Li5 & 0.750 & 0.755 & 0.254\\  
        & &  Li6 & 0.250 & 0.255 & 0.246\\  
        & &  Li7 & 0.250 & 0.245 & 0.746\\  
        & &  Li8 & 0.750 & 0.745 & 0.754\\ 
\hline
\end{tabular}
\caption{
{\bf Supplementary Table SI}. DFT relaxed crystal structure and atomic coordinates of different Weyl orbital semimetals.}
\label{tab-structure}
\end{table*}
\end{widetext}

\section{Cohesive energy calculation}
Cohesive energy of a composition, M=A$_x$B$_y$C$_z$, is defined as 
\begin{equation}\label{eq-cohesive}
E_{\mathrm {coh}}= E_{\mathrm M} - xE_{\mathrm A} - yE_{\mathrm B} - zE_{\mathrm C}.
\end{equation}
E${_\mathrm M}$ is the total energy of the primitive cell of bulk M, while E$_{\mathrm A}$ and E$_B$ and E$_C$ are the total energy per atoms of A, B, and C species, respectively, in their bulk form.  x, y, and z are the numbers of A, B and C atoms, respectively, assembled in the primitive cell of M.  In case of a binary material M=A$_x$B$_y$ the last term in Eq~(\ref{eq-cohesive}) is omitted. Cohesive energy of considered materials are listed in supplementary Table~SII. 

\begin{table}[here]
\centering
\begin{tabular}{c c}\hline
Materials & E$_\mathrm{coh}$ (eV)\\
\hline \hline 
V$_3$S$_4$ & -9.27 \\
B1Li & -0.38 \\
PbO$_2$ & -2.48 \\
NiTi$_3$S$_6$ &-12.30\\
\hline
\end{tabular}
\caption{{\bf Supplementary Table SII}. Theoretically calculated cohesive energy for different Weyl orbital semimetal classes.}
\label{tab-cohesive}
\end{table}


\end{document}